\documentclass[preprintnumbers,prd,twocolumn,nofootinbib]{revtex4-1}
\usepackage{amssymb,amsmath}
\usepackage{graphicx,color}
\usepackage{amscd}

\newcommand{\non}{\nonumber\\}

\newcommand{\tr}{{\rm tr}}
\newcommand{\UA}{U(1)$_{\rm A}$\ }
\newcommand{\m}{ \mathcal{M} }
\newcommand{\simgt}{\lower.5ex\hbox{$\; \buildrel > \over \sim \;$}}
\newcommand{\simlt}{\lower.5ex\hbox{$\; \buildrel < \over \sim \;$}}

\begin{document}
\title{
Random matrix model at nonzero chemical potentials with anomaly effects
} 

\author{H.~Fujii$^a$ and T.~Sano$^{a,b}$}
\address{%
$^a$Institute of Physics, The University of Tokyo, \\
Tokyo 153-8902, Japan\\ 
$^b$Department of Physics, The University of Tokyo, \\
Tokyo 113-0033, Japan\\ 
}

\date{\today}

\begin{abstract}
Phase diagram of the chiral random matrix model with \UA breaking term is studied
with the quark chemical potentials varied independently at zero temperature, 
by taking the chiral and meson condensates as the order parameters.
Although, without the \UA breaking term,
chiral transition of each flavor
can happen separately responding to its chemical potential,
the \UA breaking terms mix the chiral condensates and correlate
the phase transitions.
In the three flavor case, we find that
there are mixings between the meson and chiral condensates
due to the \UA anomaly, 
which makes the meson condensed phase more stable.
Increasing the hypercharge chemical potential ($\mu_Y$) with the isospin and
quark chemical potentials ($\mu_I$, $\mu_q$) kept small,
we observe that the kaon condensed phase becomes the ground state
and  at the larger $\mu_Y$ the pion condensed phase appears unexpectedly,
which is caused by the competition between the chiral restoration and the meson condensation.
The similar happens when $\mu_Y$ and $\mu_I$ are exchanged, and the kaon condensed phase
becomes the ground state at larger $\mu_I$ below the full chiral restoration. 
\end{abstract}

\maketitle

\section{Introduction}

The phase diagram of QCD at finite density
has long been challenged from the perspectives on
the fundamental properties of dense matter realized
in the core of a compact star and relativistic heavy ion collisions. 
The chiral symmetry breaking in light quark sector is 
one of the crucial features in low-energy QCD, which 
should be reflected in any effective model.
In the massless quark limit with flavor number $N_f=$ 2 or 3,
chiral symmetry breaking gives rise to continuously degenerate vacua
and to emergence of the massless Nambu-Goldstone bosons 
as the most important degrees at low energies.
In reality, nonzero u- and d-quark masses and heavier s-quark mass
lift this degeneracy and explain the observed pseudo-scalar meson spectrum.

A system at finite quark number densities is characterized 
with the quark number chemical potentials 
$\mu_f$ ($f=$ u, d, s) in the grand canonical description. 
The simplest situation at finite density may be 
the case with equal chemical potentials $\mu=\mu_u=\mu_d=\mu_s$  for all flavors,
where the s-quark density may be suppressed compared with the u- and d-quarks
due to the mass difference.
However, the quark chemical potentials $\mu_f$ are generally unequal
in various physical situations.
There are more protons than neutrons in nuclei and also in neutron stars
because of the electric charge of protons.
In the core of compact stars in beta equilibrium,
the heavy mass of the s-quark may be overcome by the large electric chemical potential,
which may result in the appearance of s-quark degrees of freedom.
In ultra-relativistic heavy-ion collisions the total strangeness number is constrained to zero.

The phase structure at finite chemical potentials has very rich theoretical
possibilities\cite{CMP_QCD,FHreview,PTPS_review,Alford:2007xm}. 
Historically, in dense nuclear matter, 
nucleon superfluidity, pion and kaon condensations, and hyperon mixture
were studied based on empirical nucleon interactions
and chiral perturbation theory\cite{PTPS_review}.
More recently, color superconductivity was extensively investigated and
the kaon condensed phase acquired renewed interests there\cite{Alford:2007xm}.
Possibility of inhomogeneous phases is also discussed recently\footnote{%
The p-wave pion condensed phase too accompanies
inhomogeneity\cite{PTPS_review}.}~\cite{InhomogeneousPhases}.

Another motivation to consider the phase diagram with unequal chemical potentials is 
that one can directly examine in lattice QCD simulations 
the $N_f=2$ case 
with finite isospin and zero quark chemical potentials,
$\mu_I \equiv (\mu_u - \mu_d)/2 \ne 0$ and 
$\mu_q \equiv (\mu_u + \mu_d)/2 =0$,
where the measure for the importance sampling is real. 
We note that it may be possible to simulate the $N_f=3$ case 
with $\mu_I \neq 0$ and $\mu_q=\mu_s=0$, at least in principle.

The ground state at finite isospin chemical potential 
but at zero quark number density was studied 
with the chiral Lagrangian, which revealed 
that the pion condensate appears 
once $\mu_I$ exceeds half the pion mass $m_\pi/2$ \cite{Son:2000xc}. 
Indeed, in the chiral limit, the degenerate ground state is
completely chiral-rotated to the pion condensed state at
infinitesimal external field $\mu_I \ne 0$.
Adding a finite quark mass $m_q \ne 0$, 
which makes the pion massive,
we have a competition between the two alignments; 
$\left < \bar q q\right >$ and $\left < \bar u \gamma_5 d \right >$
directions for $m_q$ and $\mu_I$, respectively. 
In the three-flavor case,
kaon condensation appears once the
hypercharge chemical potential 
$\mu_Y$ exceeds the threshold given by $m_K$
with $\mu_I = 0$\cite{Kogut:2001id}.

The chiral random matrix (ChRM) model is one of the models which share
the chiral symmetry with QCD. Thus it will serve as a useful model
providing qualitative features of the QCD phase diagram.  
The aim of this paper is to explore the phase structure of the (ChRM) model, 
including the possibilities of the meson condensations,
as a function of the quark chemical potentials $\mu_f$ ($f$=u, d, s) 
at zero temperature.


In the previous work of the ChRM model with two flavors 
done in Ref.~\cite{Klein:2003fy} 
it was shown that the pion condensed phase appears 
at finite $\mu_I$ above a critical value $\mu_{Ic}$
as suggested by \cite{Son:2000xc},
whereas it eventually disappears as $\mu_I$ is further increased,
which is understood as the chiral restoration at high density. 
It was also indicated that with small nonzero $\mu_I$ 
the chiral restoration occurs in two steps:
as $\mu_q$ is increased with (e.g.) negative $\mu_I<0$ fixed,
the $\left < \bar d d \right >$ condensate discontinuously decreases first,
and then the $\left < \bar u u \right >$ at slightly larger $\mu_q$.
The two- (or three-) step transition along the $\mu_q$ axis
with finite $\mu_I$ and/or $\mu_Y$ is also observed 
in the three flavor case of the ChRM model\cite{Arai:2008zz,Arai:2009zzb} and 
in the two and three flavor cases of 
the Nambu--Jona-Lasinio (NJL) model
\cite{Toublan:2003tt,Frank:2003ve,Barducci:2004tt,Barducci:2004nc}.

This two-step transition happens evidently because 
these models do not include flavor mixing interactions
and then each quark sector independently responds 
to each quark chemical potential $\mu_f$.
The \UA anomaly in QCD is known to induce the flavor mixing
and tends to unify the two chiral transitions into one. 
This is demonstrated in the NJL model with \UA breaking effective interactions,
which results in the usual phase diagram  with a single transition\cite{Frank:2003ve}.

The \UA anomaly effect is indispensable to build a 
low-energy effective model of hadrons in medium as well as
in vacuum\cite{KobaM,'tHooft:1986nc,Pisarski:1983ms}.
Although in a conventional ChRM model 
the \UA anomaly effect is not treated appropriately
so as to affect the nature of the chiral transition,
recently we have succeeded in constructing the model 
in which the \UA breaking term generates the flavor-number dependence of the
transition order\cite{Sano:2009wd}. 
This model in the massless quark limit shows the second-order phase 
transition for $N_f=2$, whereas the first-order one for $N_f=3$.
We explored the phase diagram of the model in the $T$--$\mu_q$--$m_q$ space
with $\left < \bar q_f q_f \right>$ as the order parameter\cite{Fujii:2009fm}.

It seems now quite intriguing to investigate the phase diagram
allowing the pion and kaon condensates in the ChRM model with $N_f=$2 and 3
as a function of the chemical potentials $\mu_f$,
with emphasis on the role of the \UA anomaly.
Similar studies were previously performed
employing the NJL model with \UA anomaly effects\cite{Frank:2003ve, He:2005nk, Barducci:2005ut}.
In Ref.~\cite{Frank:2003ve} the phase structure for $N_f=2$ is studied  
only in small $\mu_I$ region below the threshold for the pion condensation, 
and the study is extended  by including the pion condensed phase in Ref.~\cite{He:2005nk}.
Ref.~\cite{Barducci:2005ut} deals with the $N_f=3$ case restricted
at $\mu_I\neq 0$ and $\mu_Y= 0$, where the possibility of 
kaon condensation is ignored.

This paper is organized as follows. 
After a review of the QCD phase structure with finite chemical
potentials in Sec.~2, 
we introduce the ChRM model with anomaly effects in Sec.~3. 
The model phase structure including the possibility of the 
mesonic condensation
is investigated analytically and numerically in Sec.~4 and 5 
for $N_f=$ 2 and 3, respectively.
Sec.~6 is devoted to a summary.

\section{QCD with quark chemical potentials}

In this section we briefly review the symmetry breaking pattern of QCD with 
nonzero quark chemical potentials.
The QCD partition function with $N_f$ quark flavors 
at finite temperature and chemical potential
is regarded as the average of the Dirac operators with 
non-Abelian gauge field action $S_{\rm YM}$:
\begin{align}
Z_{\rm QCD}
=
\int {\cal D} A 
\prod_f^{N_f}
\det (D(\mu_f) + m_f)
e^{-S_{\rm YM}}
,
\label{e:QCD}
\end{align}
where $D(\mu_f)=\gamma_\nu (\partial_\nu - ig A_\nu + B_\nu \mu_{f})$ 
is the (Euclidean) Dirac operator for a flavor $f$
and $m_f$ is the mass. 
The quark number chemical potential $\mu_f$ is introduced independently 
for each quark flavor $f$ with a constant Lorentz
vector $B_\nu =(0,0,0,1)$.

In the massless limit $m_f=0$ with zero chemical potential
$\mu_f=0$, the classical QCD Lagrangian
possesses U$(N_f)_{\rm L}\times$U$(N_f)_{\rm R}$ chiral symmetry, which
is broken down to U(1)$_{\rm B}\times$SU$(N_f)_{\rm L}\times$SU$(N_f)_{\rm R}$ at
the quantum level due to the axial anomaly.
The U(1)$_B$ invariance results in the quark number conservation.
Non-perturbative QCD dynamics is believed to break this symmetry
spontaneously to U(1)$_{\rm B}\times$SU$(N_f)_{\rm V}$, generating a large part of the
nucleon mass and the light pions as the Nambu-Goldstone (NG) boson.
In reality, nonzero quark mass term acts as an external field
to select the ordinary vacuum state with nonzero $\left < \bar q q \right >$ condensate
out of the nearly degenerate vacua.

The chiral symmetry of QCD is intact at 
finite quark number chemical potential as far as
it is flavor-singlet, which is usually assumed in the context of 
finite density QCD.
Once the flavor-nonsinglet potential is set nonzero,
it gives a stress on the flavor symmetric ground state 
in addition to the quark mass term.
Symmetry breaking pattern in this situation has been studied most
conveniently with the chiral Lagrangian\cite{Son:2000xc,Kogut:2001id}.

Let us consider the $N_f=2$ case with the small u- and d-quark masses
$m_u=m_d$.
At zero chemical potential there appear nearly degenerate vacua
connected with each other 
by the SU(2)$\times$SU(2)$\simeq$O(4) transformation, and the true
ground state  aligns to the direction of the quark mass term
with SU(2)$_{\rm V}$ invariance. 
When the isospin chemical potential
$\mu_I=(\mu_u - \mu_d)/2$  is applied, this invariance explicitly
breaks down to U(1) $\in$ SU(2)$_{\rm V}$ around the 3rd isospin axis. 
The mass $m$ and the potential $\mu_I$ compete to fix the ground state.
For $\mu_I > m_\pi/2$, a new class of the degenerate vacua appears
with nonzero pion field and the ground state spontaneously
breaks the U(1) symmetry, which is accompanied 
by the appearance of an NG boson. This
symmetry breaking pattern is summarized as follows: 
\begin{align}
\begin{array}{ccl}
U(1)_{\rm B}\times SU(2)_{\rm V} & \xrightarrow{~\mu_I \ne 0~}  &
U(1)_{\rm B} \times U(1) \\
                     & \xrightarrow{~\pi^c~{\rm cond}} &
U(1)_{\rm B} \; .
\end{array}
\end{align}

For $N_f=3$, we consider the 2+1 flavor case where the quark masses 
$m_u=m_d \neq m_s$ and the chemical potentials 
$\mu_u \ne \mu_d\neq \mu_s$ in general.
In this case, all flavor symmetry is broken explicitly
leaving only U(1)$\times$U(1)$\times$U(1) invariance
which represents the conservation of each flavor.
Once (e.g.,) $\bar d \gamma_5 s$ meson condensate is formed
in addition to the nonzero chiral condensates, 
the symmetry is spontaneously broken down to  
U(1)$\times$U(1)$_{\rm ds}$,
where U(1)$_{\rm ds}$ leaves  
the condensate $\left < \bar d \gamma_5 s \right > $ invariant. 
We also find one NG mode corresponding to the spontaneous breaking of U(1)
\footnote{%
It is known that the numbers of the broken generators and the NG bosons
can differ in some cases, e.g., feromagnetism, 
kaon condensed phase with $\mu_I=0$, etc.\cite{Schafer:2001bq,Nambu}.
}.
The symmetry breaking pattern is
\begin{align}
U(1)_{\rm B}\times SU(3)_{\rm V}
& \xrightarrow{~\mu_f \ne 0~}  
U(1) \times U(1) \times U(1)
\non
&  \xrightarrow{~K~{\rm cond}}
U(1) \times U(1)_{\rm ds}
.
\end{align}

\section{Random matrix model}

The chiral random matrix model\cite{ChRM,Halasz:1998qr} is 
constructed based on the idea that the spontaneous breaking of the
chiral symmetry is dominated by the low-lying Dirac eigenmodes,
as is manifest in the Banks-Casher relation between the chiral condensate
and the spectral density near zero\cite{Banks:1979yr}.
The Dirac operator $D$ is then truncated 
to be a matrix within the restricted space 
spanned by the (quasi-) zero modes, and the matrix elements
are treated as random variables. 
Explicitly, in the representation where 
$\gamma_5={\rm diag}(1, -1)$, we have
\begin{align}
D
=
\left(\begin{matrix}
     0       & {\rm i}R+C	\\
{\rm i}R^\dagger +C^T &   0
\end{matrix}\right),
\label{e:Dirac} 
\end{align}
where $R$ is a rectangular complex random matrix 
and $C$ is a non-random matrix responsible for 
the effects of temperature and chemical potentials.  
Note that the block structure follows from the 
chiral symmetry $\{\gamma_5, D\}=0$.

When the matrix $R$ is not square but rectangular, the Dirac operator $D$
has $|\nu|$ exact zero eigenvalues with $|\nu|$ being the difference
between the numbers of the rows and the columns of $R$. 
This is the realization of the index theorem for the topological
charge and the number of exact zero modes in ChRM model.
Hence, in order to include the anomaly effects, we need to deal with
the non-square matrix properly. 
To this end, we categorize the zero modes into two species, 
the near-zero modes and the topological zero modes\cite{Sano:2009wd}. 
The former are assumed to be 
$N$ left- and $N$ right-handed low-lying modes
generated in gluon dynamics, while
the latter are interpreted as the modes 
each of which is localized near one of $N_+$ instantons or
$N_-$ anti-instantons in a gauge field configuration. 
Then the size of the matrix $R$ is taken as $(N+N_+)\times(N+N_-)$
and the topological charge of the configuration 
is $\nu = N_+ - N_-$. 
We take the number of the near-zero modes $2N$ and the 
mean number of the topological zero modes $\left < N_\pm \right >$
to be proportional to the four volume $V$ of the system.

Concerning the matrix $C$ which represents the medium
effects, we introduce the effective temperature $T$ and chemical
potential $\mu_f$ for a flavor $f$ in the near-zero mode sector,
while they are set to zero in the topological zero mode sector\cite{Halasz:1998qr,Sano:2009wd}:
\begin{align}
C_f=
\left(\begin{matrix}
(\mu_f + {\rm i}T) \mathbf{1}_{N/2} & 0                         & 0\\
     0                     &(\mu_f - {\rm i}T)\mathbf{1}_{N/2}  & 0\\
     0                     & 0                        & 0
\end{matrix}\right)
\; .
\label{eq:matter}
\end{align}
In the near-zero mode sector, 
$\mu_f \pm {\rm i} T$ may be interpreted as the contributions from 
the two lowest Matsubara frequencies.
The Dirac operator $D$ with nonzero chemical potential $\mu_f \neq 0$
is no longer anti-hermitian, though
the partition function is invariant under 
$\mu_f \leftrightarrow -\mu_f$.
We stress here the fact that the absence of the medium effects 
in the topological zero mode sector resolves 
the unphysical suppression of the topological susceptibility
of the original ChRM model (see discussion in \cite{OHTANI:2008zz,Lehner:2009xr}).
This form of $C_f$ may be understood physically
as the fact that a topological zero mode localized near
an (anti-) instanton is rather insensitive to the medium effects.

Using the Dirac operator (\ref{e:Dirac}), 
we define the partition function of the ChRM model 
with fixed $N_+$ and $N_-$ as
\begin{align}
Z_{N_+, N_-}=\int dR \; {\rm e}^{-N \Sigma^2 {\rm tr}RR^\dagger}
		 \prod_{f=1}^{N_f} \det(D(\mu_f)+m_f) 
\;
\label{eq:fixedia}
\end{align}
with a Gaussian weight for the matrix $R$.
The typical scale of the chiral symmetry breaking is fixed by the parameter $\Sigma$. 
The complete partition function is obtained by summing over 
the numbers of topological zero modes $N_+$ and $N_-$ 
with a distribution function $P(N_\pm)$ as
\begin{align}
Z^{\rm RM}
=
\sum_{N_+, N_-} P(N_+)P(N_-) Z_{N_+, N_-}
\; .
\label{eq:total}
\end{align}
The Poisson distribution would be appropriate
for $P(N_\pm)$ in a dilute instanton system\cite{'tHooft:1986nc}. 
Adopting the Poisson distribution, however,
one finds that the effective potential becomes unbounded 
from below\cite{JanikNZ97}.
Instead, we choose $P(N_\pm)$ to be a binomial distribution\cite{Sano:2009wd}, 
\begin{align}
P(N_\pm)=
\left(\begin{matrix}
\gamma N \\ 
N_{\pm}
\end{matrix}\right) 
\; 
p^{N_\pm} (1-p)^{\gamma N-N_\pm}
,
\label{eq:bino}
\end{align}
with parameters $\gamma$ and $p$.
The binomial distribution models the situation that,
if we divide the four-volume $V$ into $\gamma N$ cells,
an instanton appears in one of the cells with the probability $p$,
barring double occupancy.
In other words, we introduced a repulsive interaction 
among instantons.
This modification results in a bounded effective potential with  
a stable ground state\cite{Sano:2009wd}.

After the standard bosonization manipulation, we
can express the partition function at $T=0$ as
\begin{align}
Z^{\rm RM}(\m,\; \hat \mu)
=&
\int dS \; {\rm e}^{-2N \Omega }
\label{eq:zrm}
\end{align}
with the effective potential
\begin{align}
\Omega(S;\; \m,\; \hat \mu)
&=
\frac{\Sigma^2}{2} \text{tr} S^\dagger S
-\frac{1}{2}
\log \det
\left [ 
\begin{matrix}
S+ \m        &  \hat \mu \\
\hat \mu     & S^\dagger+ \m^\dagger 
\end{matrix}
\right ] 
\non 
&
-\frac{\gamma}{2} \log
\left | \alpha \det (S         + \m        ) +1 \right |^2
\; ,
\label{e:pot}
\end{align}
where the bosonic field $S \in {\Bbb C}^{N_f \times N_f}$ is 
the order parameter matrix 
corresponding to the bi-fermion field
$S_{fg}\sim \bar \psi_{R}^f \psi_{L}^g$,
and $\m={\rm diag}(m_{\rm u}, m_{\rm d}, \dots, m_{N_f})$ and 
$\hat \mu ={\rm diag}(\mu_{\rm u}, \mu_{\rm d}, \dots, \mu_{N_f})$ are 
the mass and the chemical potential matrices, respectively. 
The term involving the parameters $\gamma$ and $\alpha = p/(1-p)$ 
breaks the \UA symmetry
$S \to S e^{i\theta}$ even in the $\m=0$ limit\cite{KobaM,'tHooft:1986nc}.
In the thermodynamic limit, 
$N \to \infty$, the ground state is found as a solution of
the saddle point equation,
$\partial \Omega/ \partial S_{fg}=0$. 
Using the solution $\bar S$, the chiral condensate can be computed 
for a flavor $f$ as
\begin{align}
\left< \bar \psi_f \psi_f \right> 
= 
\frac{\partial }{\partial m_f}\Omega(\bar S;{\cal M},\hat \mu)
=
-\Sigma^2 \bar S_{ff}
\; .
\end{align}
By generalizing the mass matrix to the source matrix
${\cal M}=(s^a + ip^a )\lambda^a/\sqrt{2}$ with the generators
$\lambda^a$ normalized $\tr(\lambda^a \lambda^b)=2\delta^{ab}$,
other scalar (chiral) and pseudoscalar (meson) condensates are evaluated
by differentiating in $s^a$ and $p^a$, respectively.


In the preceding works\cite{Sano:2009wd,Fujii:2009fm},  
we have studied the ground state of this model for $N_f=$ 2 and 3 at
finite temperature $T$ and equal chemical potential $\mu$.
Regarding the anomaly parameters $\alpha$ and $\gamma$, 
we have found that, for the large anomaly parameters, 
$\alpha \gamma \simgt \Sigma^2$, 
this model does not show chiral restoration, 
and therefore we should use the anomaly parameters 
in the region which allows the chiral phase transition. 
In the chiral limit the chiral phase transition at finite $T$ and zero
$\mu$ is found to be second (first) order for $N_f=$2 (3). This
flavor-number dependence comes from the \UA anomaly term.
Extending to the $\mu\neq 0$ case, 
we find that the transition on the $T$--$\mu$ plane 
changes from the second order
to the first order at a tri-critical point
as $\mu$ is increased in the $N_f=2$ case. 
For $N_f=3$ the first-order phase boundary separates the $T$--$\mu$
plane into two regions. With increasing the quark mass,
the thermal transition gets weakened and eventually turns to a smooth
crossover, while the transition at larger $\mu$ than a critical point
remains of first order.
Then the $T$--$\mu$ phase diagram becomes similar to the
$N_f$=2 case with small quark masses. 
For more detailed discussions, see Refs.~\cite{Sano:2009wd,Fujii:2009fm}.

\section{Phase diagram: $N_f=2$ case}

\subsection{Effective potential}

We study the situation where $\mu_u\neq \mu_d$ with 
the degenerated quark masses $m_u=m_d \equiv m$. 
It is convenient to define the (averaged) quark chemical potential $\mu_q$ 
and the isospin chemical potential $\mu_I$ as 
\begin{align}
\mu_q &= \frac{1}{2}(\mu_u + \mu_d), \\
\mu_I &= \frac{1}{2}(\mu_u - \mu_d)
.
\end{align}
The order parameter matrix can be parametrized as
$S= \lambda_a  (\phi_a + {\rm i}\rho_a)$ with the U(2) generators 
$\lambda_a$ $(a=0,1,2,3)$.
The usual ground state breaks the chiral symmetry spontaneously,
having nonzero chiral condensate $\phi_{0}$. 
At finite $\mu_I$ the pion condensed phase
where $\rho_{1,2}$ are nonzero may be favored.
Therefore, we adopt here the following Ansatz:
\begin{align}
S=
\left(
\begin{matrix}
\phi_u	&	{\rm i}\rho_1 + \rho_2\\
{\rm i}\rho_1 - \rho_2 &	\phi_d	
\end{matrix}
\right)
\end{align}
with real order parameters $\phi_u$, $\phi_d$ and $\rho_{1,2}$. 
Notice that nonzero $\mu_I$ explicitly breaks
the SU(2) isospin invariance down to U(1) invariance generated by $\lambda_3$,
and then $\phi_u \neq \phi_d$ in general. 
Substituting this form into the effective potential (\ref{e:pot}), 
we obtain 
\begin{align}
\Omega=&
\frac{\Sigma^2}{2}(\phi_u^2 + \phi_d^2 + 2|\rho|^2) 
\non
&-\frac{1}{2}
\log \left[  (\sigma_u + \mu_u)(\sigma_d - \mu_d)+|\rho|^2 \right ]
\non
&-\frac{1}{2}
\log \left [ (\sigma_u - \mu_u)(\sigma_d + \mu_d) + |\rho|^2 \right ]
\non
&
-\frac{\gamma}{2}
\log \left[ \alpha (\sigma_u \sigma_d + |\rho|^2) +1 \right]^2
\; ,
\end{align}
where $\sigma_f=\phi_f + m$. 
This model reduces to the one studied in Ref.~\cite{Klein:2003fy}
when the anomaly term is neglected, $\gamma \alpha =0$, as it should.
Note that the potential depends on
$\rho_1$ and $\rho_2$ only through  $|\rho|^2=\rho_1^2 + \rho_2^2$
due to the residual U(1) symmetry. 
Hereafter, we shall arbitrarily choose the meson condensate such that
$\rho_1=0$ and $\rho_2=\rho$, which breaks the U(1) invariance.
Nonzero $\rho
= 
(\left< \bar u \gamma_5 d \right > - \left <\bar d \gamma_5 u \right>)/2$ 
signals the pion condensation.

The saddle point equations with respect to $\phi_u$,
$\phi_d$ and $\rho$ respectively yield
\begin{widetext}
\begin{align}
\frac{\partial \Omega}{\partial \phi_u}=
&\Sigma^2 \phi_u
-\frac{1}{2}\left[ 
\frac{\sigma_d - \mu_d}{(\sigma_u + \mu_u)(\sigma_d - \mu_d)+\rho^2}
+
\frac{\sigma_d + \mu_d}{(\sigma_u - \mu_u)(\sigma_d + \mu_d)+\rho^2}
\right]
-\frac{\gamma \alpha \sigma_d}{\alpha (\sigma_u \sigma_d +\rho^2)+1}
=
0, 
\label{e:gapu}
\\
\frac{\partial \Omega}{\partial \phi_d}=
&
\Sigma^2 \phi_d
-\frac{1}{2}\left[ 
\frac{\sigma_u + \mu_u}{(\sigma_u + \mu_u)(\sigma_d - \mu_d)+\rho^2}
+
\frac{\sigma_u - \mu_u}{(\sigma_u - \mu_u)(\sigma_d + \mu_d)+\rho^2}
\right]
-\frac{\gamma \alpha \sigma_u}{\alpha (\sigma_u \sigma_d +\rho^2)+1}
=
0
, 
\label{e:gapd}
\\
\frac{\partial \Omega}{\partial \rho}=
&
2 \Sigma^2 \rho
-\frac{1}{2}\left[ 
\frac{2\rho}{(\sigma_u + \mu_u)(\sigma_d - \mu_d)+\rho^2}
+
\frac{2\rho}{(\sigma_u - \mu_u)(\sigma_d + \mu_d)+\rho^2}
\right]
-\frac{2 \gamma \alpha \rho}{\alpha (\sigma_u \sigma_d +\rho^2)+1}
=
0. 
\label{e:gapr}
\end{align}
\end{widetext}
Let us first classify the solutions with $\rho=0$. 
Remember that when $\alpha \gamma=0$,
the chiral condensates $\phi_u$ and $\phi_d$ decouple from each other. 
Thus, in the ideal case of $m=0$, we have four solutions by making combinations of
$\phi_f=0$ and $\phi_f \neq 0$, depending on the values of $\mu_f$.
If $\alpha\gamma \neq 0$, however, 
the condensate $\phi_d$ cannot vanish exactly once $\phi_u \ne 0$
because of the flavor mixing, while the trivial solution $\phi_u=\phi_d=0$ is still intact.
Thus, in the chiral limit with $\alpha\gamma \neq 0$, we expect four types of solutions, 
(i) ordinary phase with broken chiral symmetry, 
(ii) phase with restored chiral symmetry, 
(iii) \& (iv) phases with nearly vanishing chiral condensate for only one of the flavors, u and d.
We numerically find that the transitions 
between these four phases are of first order,
and hence these four phases remain at small $m \neq 0$.

We next consider in what conditions the $\rho \neq 0$ solution appears. 
To this end, we expand the thermodynamic potential with respect to
$\rho$ at the ordinary ground state,
$\partial \Omega/\partial \phi_{u,d}=0$ and $\rho=0$:
\begin{align}
\Omega (\rho; m, \mu_f)
&=
\Omega_0(m,\mu_f) + \Omega_2 (m,\mu_f) \rho^2 + \cdots
\; .
\end{align}
Then one finds the coefficient $\Omega_2$ up to ${\cal O}(m,\mu_I^2)$ as
\begin{align}
\Omega_2 = M_\pi^2 - \frac{ 2 \mu_I^2 }{(\phi_0^2-\mu_q^2)^2}
\left  (1+ \frac{\mu_q^2}{(\phi_0^2 - \mu_q^2)^2\Sigma^2}\right )^{-1}
\; , 
\label{e:omega2}
\end{align}
where $\phi_0$ is the scalar condensate at $m=\mu_I=0$.
We see that the finite pion mass $M_\pi^2= m \Sigma^2/\phi_0$
disfavors the pion condensation, while
the finite chemical potential $\mu_I$ favors it.
From the condition $\Omega_2 =0$, 
the critical isospin chemical potential is found to be $\mu_{Ic}^2 \propto M_\pi^2$
within this leading approximation\footnote{%
Note that we use a different notation $M_\pi$ for the pion mass in this model from
the physical one $m_\pi$ 
because the physical unit in the ChRM model is not fixed.
}.

When $m=0$ {\it i.e.} $m_\pi^2=0$, infinitesimal $\mu_I$ selects out 
the ground state solution of $\phi_u=\phi_d=0$ and $\rho\neq 0$ 
among the degenerate vacua.
We find that from Eq.~(\ref{e:gapr})
the non-trivial solution satisfies the relation
\begin{align}
\rho^2
=
\mu_u \mu_d 
+ 
\left ( {\Sigma^2}-\frac{\alpha\gamma}{\alpha \rho^2 +1}\right )^{-1}
\; .
\label{e:rho}
\end{align}
For $\alpha \gamma =0$ this recovers
the solution of \cite{Klein:2003fy} 
\begin{align}
\rho_0^2
&=
\mu_u \mu_d
+\frac{1}{\Sigma^2}
=
\mu_q^2 - \mu_I^2
+\frac{1}{\Sigma^2}
\; .
\end{align}
In the large $\mu_I^2$ region, 
$\rho^2 >0$ solution disappears and the U(1) symmetry is restored.
The second-order transition line is given by a hyperbola
on the $\mu_q$--$\mu_I$ plane,
obtained by setting $\rho=0$ in Eq.~(\ref{e:rho}).
This pion condensed region is enlarged by a factor 
$(1-\alpha\gamma/\Sigma^2)^{-1}$ as compared to the case without anomaly.

Before analyzing the ground state solution numerically,
we remark here the symmetry of the effective potential (\ref{e:pot}). 
First 
the effective potential is invariant 
under the respective charge conjugations, 
$\mu_u\to -\mu_u$ and/or $\mu_d \to -\mu_d$, when $\rho=0$. 
In addition,
the effective potential (\ref{e:pot})
is invariant under $u \leftrightarrow d$.
Reflecting these invariant operations,
the phase diagram has an eightfold structure on the $\mu_u$--$\mu_d$ 
or $\mu_q$--$\mu_I$ plane. 
Nonzero condensation $\rho$ of the charged pion, however, breaks 
the respective charge conjugation symmetry leaving only the
simultaneous one
$(\mu_u, \mu_d) \to (-\mu_d, -\mu_u)$.
Together with the flavor symmetry,
$u \leftrightarrow d$,
a fourfold structure remains in the phase diagram.
On the $\mu_q$--$\mu_I$ plane, therefore, 
it is sufficient to investigate the phase diagram 
in the first quadrant, $\mu_q>0$ and $\mu_I>0$.

\subsection{Numerical result}

\begin{figure}[tb]
\begin{center}
\includegraphics[width=0.4\textwidth]{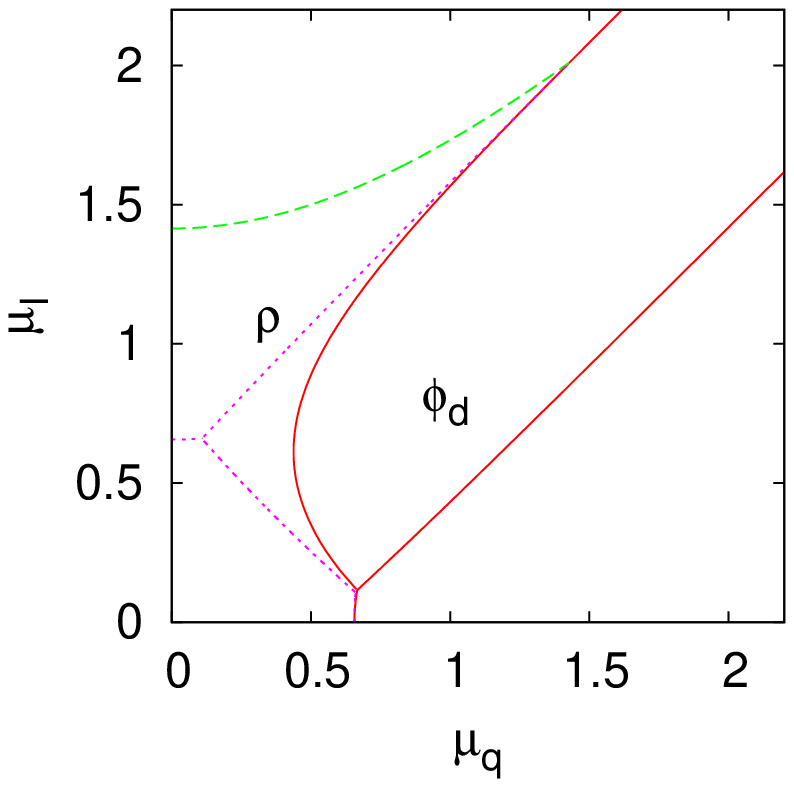}
\hfil
\includegraphics[width=0.4\textwidth]{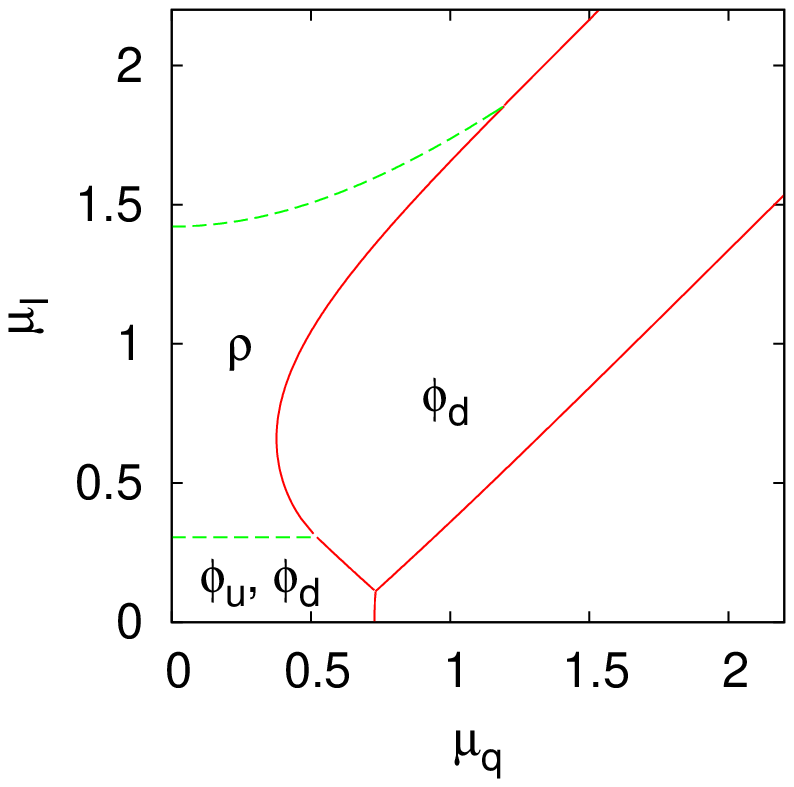}
 \caption{
Phase diagram for $N_f=2$ on $\mu_q$--$\mu_I$ plane 
in the chiral limit (top) and at $m=0.1$ (bottom).
Parameters are $\alpha=0.5$, $\gamma=1$, and $\Sigma=1$.  
Nonzero condensates in the respective regions 
are indicated with the letters, $\rho$ and $\phi_{u,d}$.
The first (second) order phase boundary is denoted in solid (dashed) lines.
For comparison, the phase boundary in the case where the meson condensation
is ignored, is denoted in dotted line showing the eightfold symmetry.
}
\label{f:Nf2}
\end{center}
\end{figure}

In the top panel of Fig.~\ref{f:Nf2}, we show the phase diagram 
on the $\mu_q$--$\mu_I$ plane with parameters,
$\alpha=0.5$, $\gamma=1$, $\Sigma=1$ and $m=0$.  
The dotted lines denotes the first-order phase transitions 
with respect to the scalar condensates $\phi_u$ and $\phi_d$ 
when the possibility of the pion condensation is neglected. 
In this restricted case we find around the origin the ordinary phase 
where $\phi_u$ and $\phi_d  \neq 0$. 
Allowing the pion condensation, we find that it completely covers the region of 
the ordinary chirally broken phase and it extends to the larger $|\mu_I|$ region, 
as shown with the solid line in Fig.~\ref{f:Nf2}.
Let us look at the small chemical potential region first. 
The ordinary phase with nonzero $\phi_u = \phi_d$ and $\rho=0$
and the pion condensed phases with $\rho \ne 0$ 
are coexisting along the line $\mu_I=\tfrac{1}{2}(\mu_u - \mu_d)=0$
in the chiral limit. 
With an infinitesimal $\mu_I$, however, 
the ordinary ground state is totally rotated away to the pion condensed phase.
This behavior was already found in the chiral sigma model\cite{Son:2000xc}. 
This is to be expected because the ChRM model 
is similar to the potential term of the chiral sigma model 
at small chemical potential $\mu_f$ and small symmetry breaking $m$\cite{Klein:2003fy}. 
Next, in the large-$\mu_f$ region in Fig.~\ref{f:Nf2},
we see that the chiral symmetry restoration occurs. 
Increasing $\mu_I$ with $\mu_q$ kept small, we find a second order phase transition 
from the pion condensed phase to the chiral restored phase. 
On the other hand, if we increase $\mu_q$ along with the fixed $\mu_I$ line, 
we find a first-order phase transition from the pion condensed phase 
to the phase where one or both of the chiral condensates $\phi_u$ and $\phi_d$
melt away, depending on the size of $\mu_I$. 
It is remarkable that the phase diagram with $\rho=0$ 
reflects the symmetry of 
$(\mu_q, \mu_I) \leftrightarrow (\mu_I, \mu_q)$ with 
$(\phi_u, \phi_d) \leftrightarrow (\phi_d, \phi_u)$, 
while in the real ground state with $\rho \neq 0$, 
this symmetry no longer exists. 
As we mentioned, the phase diagram has the symmetry of 
$(\mu_q, \mu_I) \to (\pm \mu_q, \pm \mu_I)$.

In the bottom panel of Fig.~\ref{f:Nf2} we show the result with $m = 0.1 \neq 0$. 
Generally, finite $m$ acts as an external alignment field for $\phi_u,\; \phi_d \ne 0$.
So there is a competition between the two alignment fields $m$ and $\mu_I^2$.
In a small $\mu_I$ region,
the finite $m$ wins and the ordinary phase with nonzero $\phi_u$ and $\phi_d$ 
appears pushing the pion condensed phase aside,
as shown in Fig.~\ref{f:Nf2}.
If $\mu_I$ exceeds a critical value proportional to $m_\pi$, 
the pion condensation phase appears. 
It is estimated from Eq.~(\ref{e:omega2}) 
with our model parameters 
as $\mu_{Ic}=\sqrt{\phi_0^4 M_\pi^2/2}=0.276$ at $\mu_q=0$ .
We here remark the number of the phase transitions 
in small $\mu_I$ region where $\rho=0$. 
When $\mu_q$ is increased,
the system experiences just one phase transition 
from the ordinary phase with $\phi_u>0$ and $\phi_d>0$ 
to the symmetric phase with $\phi_u\sim 0$ and $\phi_d\sim 0$ 
if $|\mu_I|$ is sufficiently small. 
If $|\mu_I|$ is large but still below the critical value, 
we have two transitions between the ordinary phase and the symmetric phase;
e.g., for $\mu_I > 0$, 
the first one is from the phase of $\phi_u>0,\; \phi_d>0$ 
to $\phi_u\sim 0,\; \phi_d> 0$, 
and the second is from $\phi_u\sim 0,\; \phi_d> 0$ 
to $\phi_u\sim 0,\; \phi_d\sim 0$. 
In the study of the ChRM model without the anomaly effect in \cite{Klein:2003fy}, 
a two-step phase transition is found. 
We find that, even with $\rho=0$, 
the anomaly term mixes the $\phi_u$ and $\phi_d$ yielding
a one-step transition. 
This mechanism is general and is found in an NJL model study\cite{Frank:2003ve}. 
We expect that this is also the case in QCD.
As a result, we have one triple point and two critical end points
\footnote{A critical end point is the point where
a critical line is truncated by meeting a first-order phase boundary.}
in the phase diagram
shown in the bottom of Fig.~\ref{f:Nf2}.
(In the $N_f =3$ case we will find more triple points and critical end points.)
Varying the anomaly parameters $\alpha$ and $\gamma$, 
we have confirmed that 
the position of the triple point moves toward the larger $\mu_q$ and $\mu_I$ region
as the strength of the flavor mixing is increased.

Finally we comment on the large $\mu_I$ region regarding the possibility of BCS-like condensate. 
At large $\mu_I$ in QCD, large Fermi seas of (e.g.) u-quark and $\bar {\rm d}$-quark are formed,
and the attraction between these quarks may form a pion-like Cooper pair\cite{Son:2000xc}.
However, since the ChRM model neglects the space-time dimensions, the physics of
the Fermi surface does not exist and the superconducting phase is tricky. 
For possible extensions of the ChRM models to deal with the diquark condensates, see Ref.~\cite{diquark}.
In our model without such an extension, 
we find a simple 
termination of the pion condensed phase at large $|\mu_I|$.

\section{Phase diagram: $N_f=3$ case}

\subsection{Effective potential}

We explore here
the phase structure in the 
2+1 flavor case with two mass parameters, 
$m_u=m_d \equiv m$ and $m_s$, 
varying three chemical potentials 
$\mu_u, \mu_d$ and $\mu_s$ independently. 
They are recast to 
the quark chemical potential $\mu_q$, 
the isospin chemical potential $\mu_I$ 
and the hypercharge chemical potential $\mu_Y$ defined respectively as
\begin{align}
\mu_q &= \frac{1}{2}(\mu_u + \mu_d), \\
\mu_I &= \frac{1}{2}(\mu_u - \mu_d), \\
\mu_Y &= \frac{1}{2}(\mu_u + \mu_d - 2\mu_s)
.
\end{align}
Note that $\mu_Y = \mu_q - \mu_s$ in our convention.
When $\mu_I=\mu_Y=0$, three quark chemical potentials are equal to $\mu_q$. 
At nonzero $\mu_I$ and $\mu_Y$, we allow the possibilities for 
the pion and kaon condensates in addition to the chiral condensates. 
We then apply the following Ansatz for the order parameter matrix, 
\begin{align}
S=
\left(
\begin{matrix}
\phi_u	&	\rho_{ud} & -\rho_{su}	\\
-\rho_{ud}	&	\phi_d & \rho_{ds}	\\
\rho_{su}&	-\rho_{ds} & \phi_s
\end{matrix}
\right), 
\label{e:ansatz_nf3}
\end{align}
where $\phi_u$, $\phi_d$, and $\phi_s$ are the chiral condensates, 
$\rho_{ud}$ is the pion condensate and 
$\rho_{su}$ and $\rho_{ds}$ are the kaon condensates. 
With this Ansatz,
we obtain the effective potential as an function of 
six order parameters, 
\begin{align}
\Omega
=
\Omega_{\rm 0}
+
\Omega_{\rm a}
,
\end{align}
where $\Omega_{\rm 0}$ is the potential of the conventional ChRM model\cite{Arai:2008zz},
\begin{widetext}
\begin{align}
\Omega_{\rm 0}
=&
\frac{\Sigma^2}{2}
\left(
	\phi_u^2 + \phi_d^2 + \phi_s^2 
	+ 2\rho_{ds}^2+2\rho_{su}^2+2\rho_{ud}^2
\right)
-\frac{1}{4}
\log \left \{
(\sigma_u^2 -\mu_u^2)(\sigma_d^2 -\mu_d^2)(\sigma_s^2 -\mu_s^2)
\right .
\non 
& +\left .
\left [ \rho_{ds}^4 (\sigma_u^2 -\mu_u^2)  
+2\rho_{ds}^2 \rho_{su}^2 (\sigma_u \sigma_d - \mu_u \mu_d)
+2 \rho_{ds}^2 (\sigma_u^2 - \mu_u^2)(\sigma_d \sigma_s - \mu_d \mu_s)
+(\mbox{cyclic perm. of u,d,s}) \right ]
\right \}^2
,
\end{align} 
\end{widetext}
and $\Omega_{\rm a}$ is the anomaly part%
\footnote{%
Consequences of the anomaly mixing between the chiral and diquark condensates 
are studied in Ref.~\cite{Hatsuda:2006ps}.
}%
,
\begin{align}
\Omega_{\rm a}
&=
-\frac{\gamma}{2}
\log \left[
\alpha
\left(
\sigma_u \sigma_d \sigma_s 
+\sigma_u \rho_{ds}^2
+\sigma_d \rho_{su}^2
+\sigma_s \rho_{ud}^2
\right)
+1
\right]^2
.
\label{e:pot_anom}
\end{align}
Because $\Omega$ is a function of $\rho_{ud}^2$, $\rho_{ds}^2$ and $\rho_{su}^2$, 
we always have a trivial solution $\rho_{fg}=0$ for the saddle point equations. 
On the other hand, once chiral and/or meson condensates become nonzero,
they act as source terms for the other chiral condensates 
owing to the anomaly term $\Omega_{\rm a}$, and therefore
$\sigma_f=0$ is no longer a solution.

We confirmed numerically in the ChRM model that 
two or more meson condensates do not appear in the ground state at the same time. 
If only one type of the meson condensate $\rho_{fg}$ is nonzero,
the $\Omega_0$ part of the potential becomes a sum of two contributions,
$
\Omega_0 = 
\Omega_0^{fg}(\rho_{fg},\phi_f,\phi_g) 
+ \Omega_0^{h}(\phi_h) 
$, 
where $\Omega_0^{fg}$ is nothing but the potential of
the two-flavor ($fg$) ChRM model without anomaly, and 
$\Omega_0^{h}$ corresponds to the single-flavor ($h$) ChRM model.
In spite of this flavor separation, 
the phase diagram of $N_f=3$ is different from that of $N_f=2$ 
because the anomaly term introduces the coupling among three flavors 
and we have a competition between the pion ($fg=$ud) and the kaon
($fg=$ds or su) condensed phases for being the ground state.
Below, we just assume that only one component of the meson condensates
becomes nonzero in the meson condensed phase 
and leave its proof as an open issue\footnote{We note that coexistence of
the $p$-wave pion and kaon condensates in nuclear matter was studied
previously\cite{Muto:2003dk}.}.

Preceding the numerical results, 
let us summarize three key points 
for qualitative understanding of the phase structure.

(I) {\it analogy with the chiral sigma model} --- 
As is shown in Ref.~\cite{Klein:2003fy},
the conventional ChRM model 
for small chemical potentials and quark masses
is equivalent to the zero momentum part of
the chiral Lagrangian.
Hence 
the phase structure of the ChRM model in the small chemical potential region
must have a similar structure as the chiral sigma model. 
It is known in the sigma model with $N_f=3$ 
that a second-order phase transition occurs 
from the ordinary chirally broken phase
to the pion condensed phase or the kaon condensed phase 
at certain finite $\mu_I$ and $\mu_Y$. 
The critical chemical potential is roughly estimated as 
$\mu_I \sim m_{\pi}/2$ and
$\mu_Y \sim m_{K}$ for the pion and kaon condensations,
respectively.
There is a competition between the pion and kaon condensed phases
for finite $\mu_I$ and $\mu_Y$,
and the phase transition between the two is found to be first-order. 
We also remark that the state with two meson condensates having 
nonzero values can not be even a meta-stable state in this analysis.

(II) {\it chiral restoration} --- 
In contrast to the chiral sigma model, the ChRM model includes 
the chiral restoration dynamics, 
which will result in a new kind of competition of two meson condensed phases 
in the region at large chemical potential.
To illustrate the situation,
let us consider the case in the chiral limit, 
where $\mu_Y > \mu_I >\mu_q =0$ 
and the $K^0$-condensed phase $\rho_{us}\neq 0$ with $\phi_d\neq 0$ 
is chosen as the ground state.
We now further increase $\mu_Y$ with $\mu_q$ and $\mu_I$ fixed. 
This kaon condensed phase will simply remain as the ground state in the chiral sigma model.
At sufficiently large $\mu_Y$, however, chiral or meson condensates involving the s-quark
become disfavored and the chiral symmetry in the s-quark sector will presumably be restored.
Then there is a possibility for the remaining u- and d-quarks to form the pion condensate $\rho_{ud}\neq 0$.
We thus find a competition between the kaon condensed
phase ($\rho_{us}\neq 0$, $\phi_d\neq 0$) and
the pion condensed phase ($\rho_{ud}\neq 0$, $\phi_s = 0$) at large $\mu_Y$. 
We will see shortly in this section that
the pion-condensed phase is indeed favored at large $\mu_Y$.
This competition exists also in the NJL model\cite{Barducci:2004nc}, 
and we expect that this mechanism is common for the models exhibiting
chiral symmetry restoration at large chemical potentials.

(III) {\it anomaly effect} --- 
Mixing of the order parameters due to the anomaly 
tends to unify chiral phase transitions
into a single one as is seen in the case of $N_f=2$. 
In the absence of the meson condensates, 
the anomaly term $\Omega_{\rm a}$ for $N_f=3$ reduces to 
$\log (\alpha \sigma_u \sigma_d \sigma_s +1)$, 
which couples three chiral condensates. 
When one of the chiral condensates melts away,
the flavor mixing among the chiral condensates becomes small.
With finite meson condensates, however, 
the anomaly term induces mixing among the chiral and the meson condensates.
For example, once the pion condensate is formed $\rho_{ud} \ne 0$ 
with rotating $\phi_u$ and $\phi_d$ away,
the mixing term  $\sigma_u \sigma_d \sigma_s $ becomes small
but the term $\phi_s \rho_{ud}^2$ in Eq.~(\ref{e:pot_anom})
generates new flavor mixing.
Because of this mixing, the chiral and pion condensates tend to vary cooperatively
as the chemical potentials change.
Furthermore, the anomaly term 
makes the ground state with nonzero $\phi_s$ and $\rho_{ud}$ condensates 
more stable as the term appears with a minus sign in the effective potential, 
which results in the extension of the meson condensed phase
compared to the case without anomaly.

\begin{figure}[tb]
\begin{center}
\includegraphics[width=0.4\textwidth]{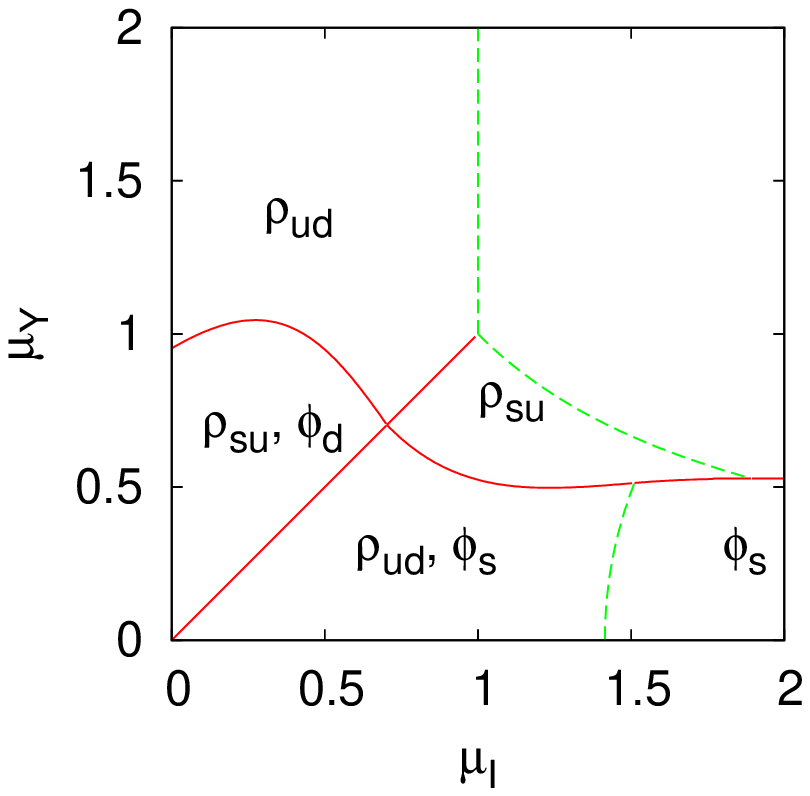}
\includegraphics[width=0.4\textwidth]{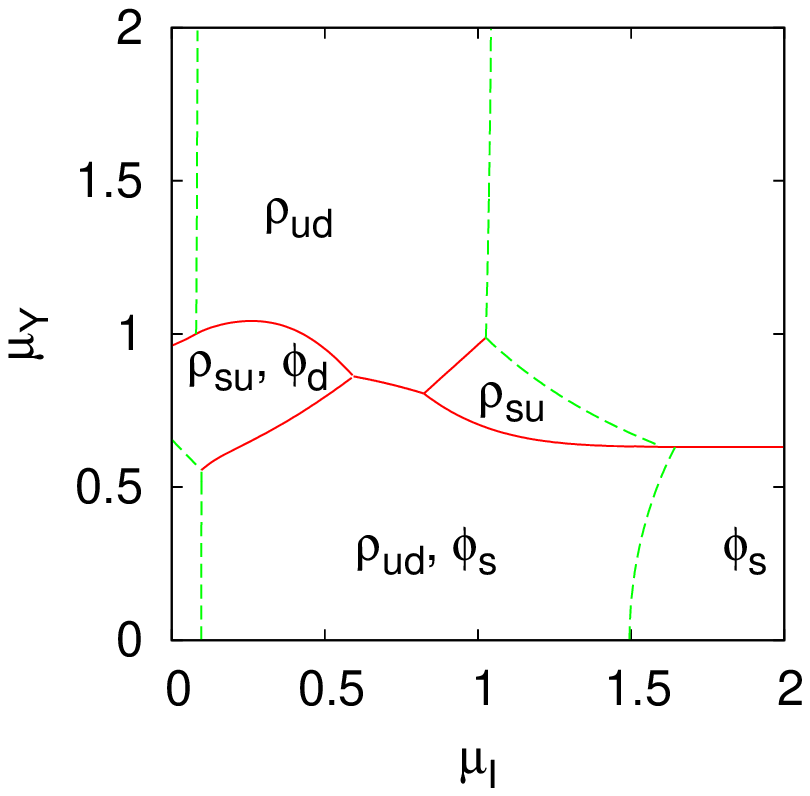}
 \caption{
Phase diagram for $N_f=3$ on $\mu_I$--$\mu_Y$ plane with $\mu_q=0$
in the chiral limit (top) and with nonzero quark masses
$m_u=m_d=0.02$, $m_s=0.1$ (bottom). 
In the bottom panel, the lower (upper) vertical narrow area represents the
phase with nonzero $\phi_u$, $\phi_d$ and $\phi_s$ ($\phi_u$ and $\phi_d$).
Other parameters and notations are the same as in Fig.~\ref{f:Nf2}. 
}
\label{f:Nf3_iy}
\end{center}
\end{figure}

\subsection{Numerical result}

We calculated the phase diagram 
with model parameters $\Sigma=1$, $\alpha=0.5$ and $\gamma=1$
in the chiral limit $m_u=m_d=m_s=0$ as well as at finite quark masses,
$m_u=m_d=0.02$ and $m_s=0.1$. 
We show the phase diagram on
the $\mu_I$--$\mu_Y$ plane with $\mu_q=0$, 
the $\mu_q$--$\mu_I$ plane with  $\mu_Y=0$, 
and the $\mu_q$--$\mu_Y$ plane with $\mu_I=0$.

\subsubsection{$\mu_I$--$\mu_Y$ plane}

We first present the phase diagram on the $\mu_I$--$\mu_Y$ plane 
with zero quark chemical potential $\mu_q=0$
in the chiral limit (top) and at finite quark masses (bottom)
in Fig.~\ref{f:Nf3_iy}. 
Note that the effective potential has the symmetry of 
$\mu_Y \leftrightarrow -\mu_Y$ and, 
moreover, $\mu_I \leftrightarrow -\mu_I$ 
if we change u and d flavors simultaneously. 
We then present the result only in the first quadrant.

Let us first focus on the case in the chiral limit. 
In the small chemical potential region, 
we find the pion and kaon condensed phases. 
These two phases are separated by 
the first-order phase transition line $\mu_I = \mu_Y$; 
the pion condensed phase appears when $\mu_I > \mu_Y$ and
otherwise the kaon condenses.
This is also found in the chiral sigma model\cite{Kogut:2001id}. 
When chemical potentials are increased, we find that 
the regions of two mesonic phases are exchanged on the diagram; 
the pion condensed phase appears when $\mu_I < \mu_Y$
and otherwise the kaon condenses.
As already explained, this peculiar behavior is triggered by 
the chiral restoration; 
these meson condensed phases at the larger chemical potentials 
are accompanied by the melting of the chiral condensate of the remaining flavor. 
When $\mu_Y$ is further increased at small $\mu_I$,
the pion condensed phase continues indefinitely because
$\rho_{ud}$ becomes insensitive to $\mu_Y$ once $\phi_s$ disappears.
On the other hand, 
$\rho_{ud}$ continuously vanishes at some point
as $\mu_I$ increases, because $\rho_{ud}$ which has the isospin charge
is affected by $\mu_I$, whereas
nonzero chiral condensate $\phi_s$ survives irrespective of $\mu_I$
for small $\mu_Y$ because it has no isospin. 

Without the anomaly term the vertical straight line of the 
second-order transition boundary seen in the large $\mu_Y$ region in Fig.~\ref{f:Nf3_iy}
would extend down to the $\mu_I$ axis.
This is because the effective potential without anomaly $\Omega_0$ becomes
the sum of two contributions, the $\phi_s$ part and the
part involving $\phi_u, \phi_d$ and $\rho_{ud}$, as mentioned before,
unless the kaon condensates have finite values. 
By comparison, we see that 
the anomaly coupling between $\phi_s$  and $\rho_{ud}$ 
makes the pion condensed phase more stable and extended to
larger $\mu_I$ region.

When quark masses are set to nonzero, 
the phase diagram receives two qualitative modifications. 
One is the appearance of the chiral condensed phase 
without any meson condensate at small chemical potentials. 
This is the same result as in the chiral sigma model. 
The other is the configuration change of the four meson condensed phases. 
We find that the pion condensed phase extends
from zero $\mu_Y$ to large $\mu_Y$, cutting the two kaon condensed phases apart.
This is because the chiral condensate $\phi_s$ is more favored 
for $m_s >m_u=m_d$ and the kaon condensed phases shrink.
In the extended pion condensed phase, we find a first-order phase transition 
in $\phi_s$, which creates a small gap in $\rho_{ud}$, too.

\begin{figure}[tb]
\begin{center}
\includegraphics[width=0.4\textwidth]{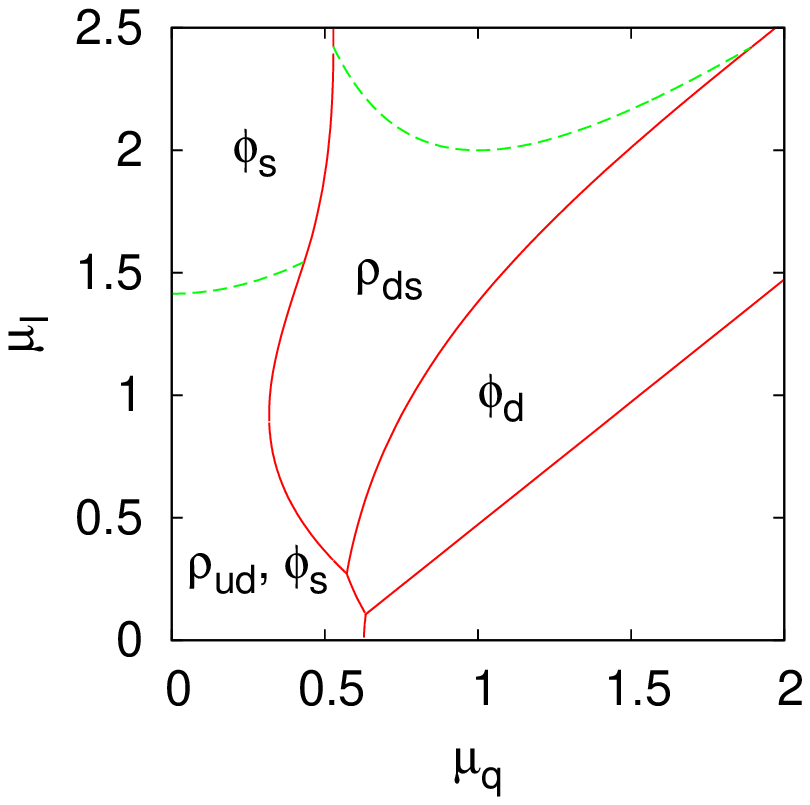}
\includegraphics[width=0.4\textwidth]{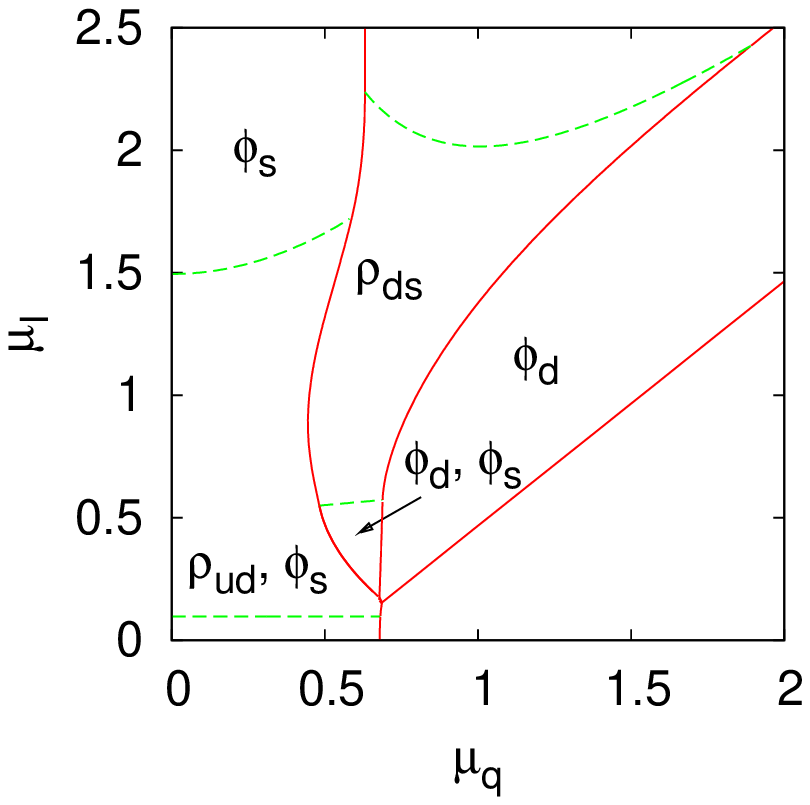}
\caption{
Phase diagram for $N_f=3$ on $\mu_q$--$\mu_I$ plane with $\mu_Y=0$
in the chiral limit (top) and with nonzero quark masses (bottom). 
In the bottom panel, the horizontal narrow area is
the phase with nonzero $\phi_u$, $\phi_d$ and $\phi_s$. 
Parameters and notations are the same as in Fig.~\ref{f:Nf3_iy}.
}
\label{f:Nf3_qi}
\end{center}
\end{figure}

\subsubsection{$\mu_q$--$\mu_I$ plane}

We next show the phase diagram on the $\mu_q$--$\mu_I$ plane with $\mu_Y=0$
in Fig.~\ref{f:Nf3_qi}.
In this case the effective potential is symmetric in 
$\mu_q \leftrightarrow -\mu_q$
and in $\mu_I \leftrightarrow -\mu_I$ with exchange of 
the u and d quarks.
We present the phase diagram again only in the first quadrant.

Let us focus on the case in the chiral limit (top panel of Fig.~\ref{f:Nf3_qi}). 
With infinitesimal $\mu_I$, 
the ground state becomes the pion condensed phase. 
Unexpectedly, however, we find that a kaon-condensed phase appears 
in the region where $\mu_u$ becomes largest among
three chemical potentials.
This phase was not predicted in the chiral Lagrangian analysis. 
In this phase, the u-quark chiral condensate melts away $\phi_u\sim 0$, 
and the remaining d- and s-quarks form the kaon condensate, $\rho_{ds}\neq 0$. 
Note that, because $m_K=0$, 
infinitesimal difference between $\mu_d$ and $\mu_s$ makes 
the kaon condensed phase more stable than the chirally broken phase without 
a meson condensate.

The pion and kaon condensed phases vanish continuously at large $\mu_I$,
respectively, to the phases with $\phi_s$ and without any condensate.
The kaon condensed phase extends to 
a larger $\mu_I$ region than the pion condensed phase. 
This may be understood from the fact that the pion condensate with isospin
charge 1 is twice as sensitive to $\mu_I$ as the kaon condensate
with isospin 1/2.

With finite quark masses we observe that
two new phases are added, as is seen 
in the bottom panel of Fig.~\ref{f:Nf3_qi}. 
The ordinary chirally broken phase appears
in the small $\mu_I$ and $\mu_q$ region due to finite $m_{\pi}$,
while finite $m_{K}$ changes lower chemical potential part of the kaon condensed region
into the chiral condensed phase with nonzero $\phi_d$ and $\phi_s$.
These two chiral condensed phases show 
the second-order phase transitions to the meson condensed phases
at larger $\mu_I$, 
which is consistent with the chiral Lagrangian analysis\cite{Kogut:2001id}.
At small $\mu_I$, we find a single transition 
from the broken to the symmetric phase along the $\mu_q$ axis,
owing to the anomaly term.

It would be instructive to compare this phase diagram to 
that of $N_f=2$ on the $\mu_q$--$\mu_I$ plane. 
Although the phase diagrams are drawn in the same chemical potential space, 
inclusion of the third quark flavor 
changes the phase diagram drastically resulting in 
a new phase with the kaon condensation. 
If we take a limit of $m_s \to \infty$ in the $N_f=3$ phase diagram,
the s-quark should decouple and the s-quark and kaon condensates disappear.
This makes the phase diagram reduce to that of the $N_f=2$ case.

\begin{figure}[tb]
\begin{center}
\includegraphics[width=0.4\textwidth]{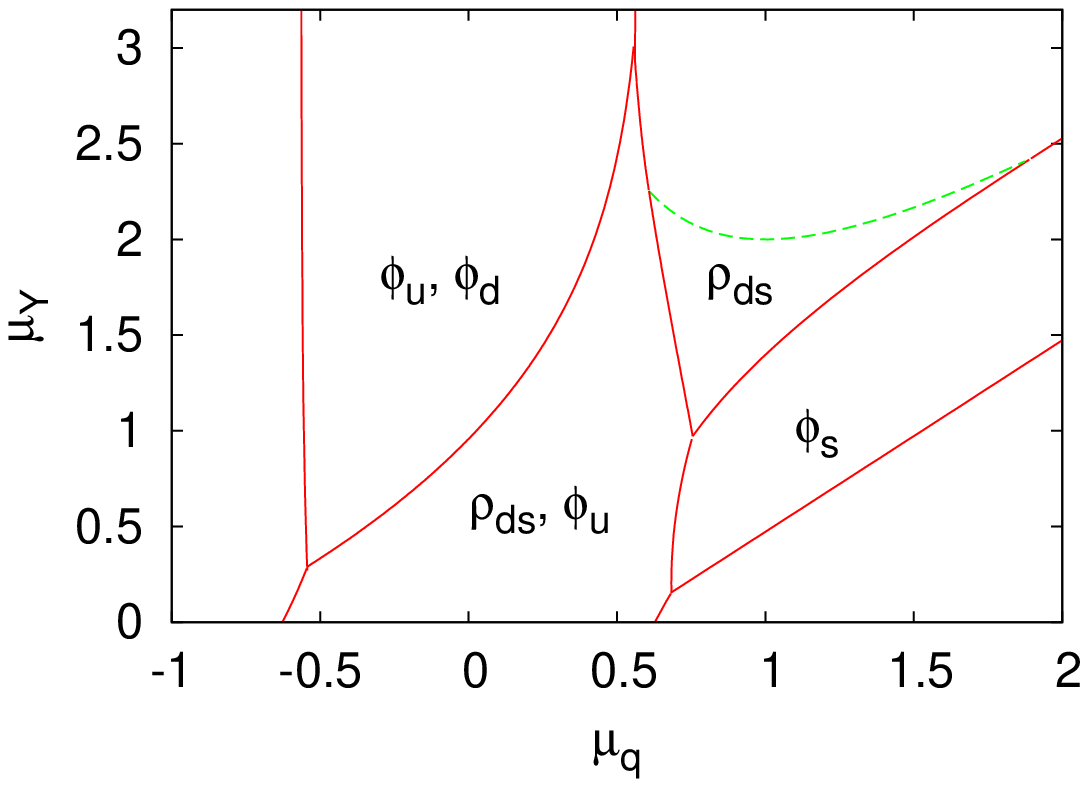}
\includegraphics[width=0.4\textwidth]{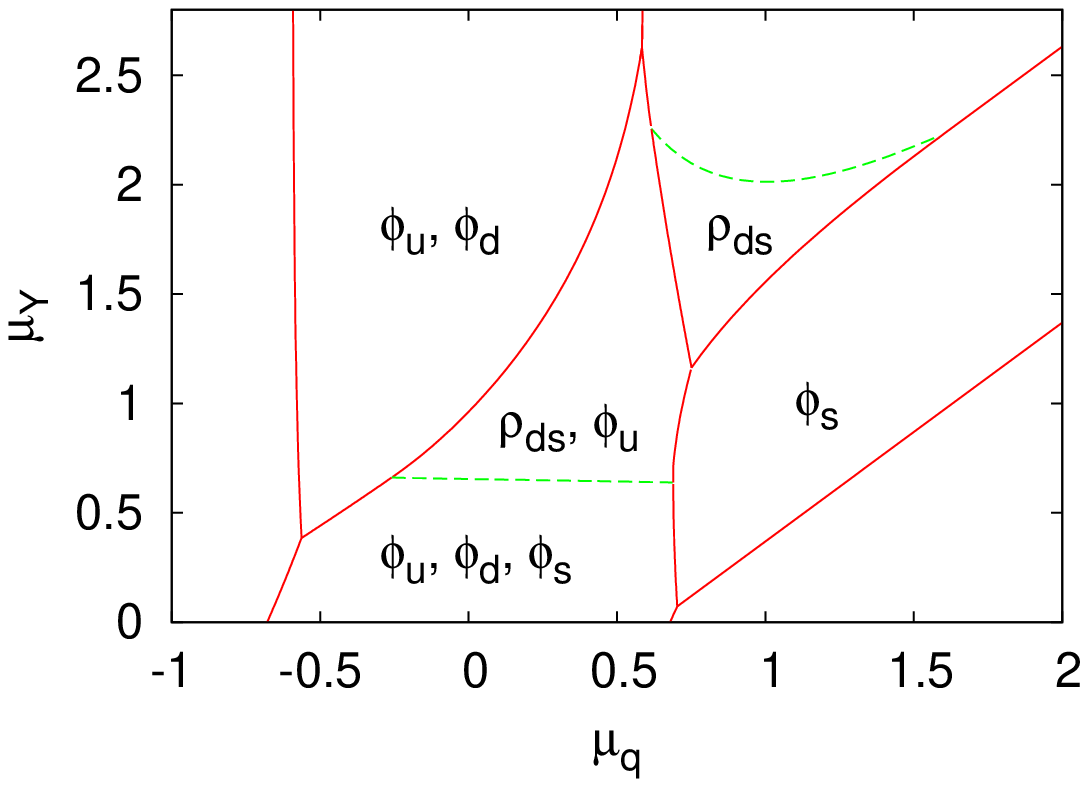}
 \caption{
Phase diagram for $N_f=3$ on $\mu_q$--$\mu_Y$ plane with $\mu_I=0$
in the chiral limit (top) and with nonzero quark masses (bottom). 
Parameters and notations are the same as in Fig.~\ref{f:Nf3_iy}.
}
\label{f:Nf3_qs}
\end{center}
\end{figure}

\subsubsection{$\mu_q$--$\mu_Y$ plane}

Finally, we address the phase diagram on the $\mu_q$--$\mu_Y$ plane with $\mu_I=0$. 
In Fig.~\ref{f:Nf3_qs} we present the phase diagram 
in the chiral limit (top panel) and with finite quark masses (bottom panel). 
The effective potential is unchanged under the simultaneous exchanges of
$\mu_q \leftrightarrow -\mu_q$ and $\mu_Y \leftrightarrow -\mu_Y$.
Hence we restrict the phase diagrams in $\mu_Y>0$ region.
The formation of the kaon condensate breaks the SU(2) isospin symmetry
spontaneously. We choose arbitrarily the $K_0$-condensed phase 
($\rho_{ds}\neq 0$) as the meson condensed ground state. 
Note that no pion condensed phase appears in this diagram
since we have set $|\mu_I| = 0 \leq m_\pi$.

In the chiral limit, 
the chiral and kaon (pion as well) condensed phases 
are degenerated at $\mu_q=\mu_Y=0$. 
With infinitesimal $\mu_Y$, 
the ground state becomes the kaon condensed phase, 
and therefore the isospin symmetry is broken spontaneously. 
Assuming the alignment by infinitesimal negative $\mu_I$, 
the chiral condensed state is completely rotated into the 
the $K^0$ condensed state where $\phi_d = \phi_s= 0$ and $\rho_{ds}\neq 0$
but the u-quark chiral condensate remains finite $\phi_u\neq 0$. 
When $\mu_q$ is increased with $\mu_Y=0$, 
two order parameters, $\rho_{ds}$ and $\phi_u$ disappear at the same time
at a critical value. 
Without the anomaly effect this phase transition would occur in two steps.

We note that there is a first-order phase boundary 
inside of the the kaon condensed phase $\rho_{ds} \ne 0$.
Across this line, $\phi_u$ jumps from a nonzero value to zero and 
$\rho_{ds}$ changes also its value discontinuously. 
At one end of this line, we find a triple point where
the two kaon condensed phases and the $\phi_s \neq 0$ phase coexist.
At the other end, we find a critical end point where
two kaon condensed phases and the symmetric phase meet.

Let us consider the finite mass effect. 
Because of the finite $m_K$, 
there appears the ordinary chirally broken phase in the 
small chemical potential region. 
The phase transition to the kaon condensed phase is of second order. 
Along the $\mu_q$ axis, we find the phase transition from  
the chiral condensed phase to the symmetric phase occurs in one step.
The threshold $\mu_Y$ for the two-step phase transition is relatively small 
compared to the corresponding value of $\mu_I$ found in the $\mu_q$--$\mu_I$ plane 
(see Fig.~\ref{f:Nf3_qi}). 

This can be understood as follows:
note that without the flavor mixing, 
the first order transitions along the $\mu_q$ axis 
would occur at different values of $\mu_q$
when the quark masses are different.
The \UA anomaly mixing acts to bind them together,
and there is a critical value for the anomaly strength 
above which the phase transitions occur in one step  
along the $\mu_q$ axis.
The larger mass difference between u, d-quarks and s-quark makes
this binding of the chiral transition more fragile against
the external field $\mu_Y$. 
This is the reason why we have a smaller threshold value of $\mu_Y$
compared to the threshold value of $\mu_I$.

With the mass parameters $m_u=m_d=0.02$ and $m_s=0.1$ used in Fig.~\ref{f:Nf3_qs}, 
the threshold value is found to be $\alpha_c= 0.279\dots $
when $\gamma =1$ is fixed. 
In Fig.~\ref{f:Nf3_qs}, $\alpha=0.5 >\alpha_c$ and we see
the one-step phase transition along the $\mu_q$ axis.


\section{Summary}

We have investigated the phase structure of the ChRM model with
2 and 3 flavors on the plane of the quark chemical potentials,
including the effects of anomaly.
Different chemical potentials for different flavors
are realized in general situations 
because of the quark mass differences and electric neutrality, 
and the chiral condensates have also different values there.
Moreover,
when the chemical potential difference, which is flavor nonsinglet, becomes large
and comparable to the pseudo-scalar meson masses,
we have a phase transition to a meson condensed phase.
We obtained a complex structure of the phase diagram with various orderings at first sight.
Nevertheless, we can understand the diagram qualitatively based on the following
three observations.

The first point is the similarity to the chiral Lagrangian analysis, 
which explains the phase structure well below the chiral restoration. 
In the chiral limit for $N_f=2$, 
the chiral condensed phase and 
the pion condensed phase are degenerated at $\mu_I=0$, and 
an infinitesimal $\mu_I$ rotates the chiral condensates into the pion condensate. 
With finite quark masses the chiral condensed phase survives up to $\mu_I\sim m_\pi/2$,
and we find a second-order phase transition to the pion condensed phase.
In the $N_f=3$ case, a similar behavior is also found for meson condensations,
but in addition we found a new competition between 
the pion and kaon condensed phases, 
bounded by a first-order phase transition.

The second point concerns the chiral restoration, 
whose effect is not included in the chiral Lagrangian
analysis. 
For $N_f=2$ and 3, 
we found that the pion and kaon condensates disappear continuously 
at large $\mu_I$ and $\mu_Y$, respectively,
while the phase transitions along the $\mu_q$ axis
are first-order. 
The chiral restoration effect explains 
the peculiar appearance of the pion and the kaon condensed phases 
in the phase diagram for $N_f=3$. 
In the chiral Lagrangian analysis, 
the pion condensation is always favored if $\mu_I > \mu_Y$. 
However when $\mu_I$ is further increased,
the chiral restoration for u-quark should occur
and the pion condensation melts away. 
Then the kaon condensation will be formed 
if the difference between $\mu_d$ and $\mu_s$ is larger 
than a certain critical value characterized by $m_K$. 
This argument also holds in the situation where $\mu_I < \mu_Y$ 
with exchanging the pion and kaon condensates.
This is a remarkable inversion of 
the pion and kaon condensation at high $\mu_I$ or $\mu_Y$. 
Note that the argument does not depend on the details of 
the ChRM model. 
The same inversion may be found in other models 
which incorporate the chiral restoration dynamics, 
and therefore one may expect this also in QCD.

The third point is about the anomaly effect. 
The most important effect is the mixing of the condensates.
For $N_f=2$ with finite quark masses, 
we find a single phase transition 
from the chiral condensed phase to the symmetric phase 
along the $\mu_q$ axis.
Without the flavor mixing,
the u- and d-quarks would show the chiral transitions independently
at finite $\mu_I$.
But these two phase transitions coalesce into a single transition
via the flavor mixing term due to the \UA anomaly. 
There is the threshold $\mu_I$ for two-step phase transition, 
which becomes large if the anomaly parameters are increased.

In the case of $N_f=3$, 
the anomaly effect induces the mixing among 
not only the chiral condensates,
but also the meson condensates,
which makes likely that the meson condensate and
the chiral condensate of the remaining flavor
show the discontinuity simultaneously as the chemical potentials are varied.
In the mean-field models without the anomaly effect, where the flavor mixing is missing,
the chiral phase transitions can happen in multi steps 
for finite $\mu_I$ and/or $\mu_Y$ even when the quarks have an equal mass.
We expect that the chiral condensates are
strongly correlated in QCD because of 
the anomaly mixing as well as the dynamics beyond the mean field, 
and 
it is unlikely to have the multi-step chiral restoration in QCD when 
$\mu_q$ is increased with $\mu_I$ and $\mu_Y$ being not too large. 
For large $\mu_I$ and/or $\mu_Y$, however, there remains a possibility 
for chiral restoration to occur in multi steps, as shown in our model study.

The ChRM model, which discards the space-time dynamics but keeps the chiral symmetry only,
is certainly a simplified model for QCD and will provide at most qualitative features 
of the QCD phase diagram.
We have studied the response of the condensates to the chemical potentials
using this ChRM model with the \UA anomaly, and found interesting interplay
between the chiral and meson condensates as well as the importance of 
the \UA anomaly effects. It will be intriguing to confirm and elaborate
the findings of this study by employing other dynamic and microscopic models for QCD 
and direct simulations.

\acknowledgements
The authors are grateful to members of Komaba nuclear theory group for their
interests in this work and encouragements.
H.F.\ acknowledges the warm hospitality extended to him by 
Institut f\"ur Kernphysik in Technische Universit\"at Darmstadt,
where this work was finalized.
This work is supported in part by Grants-in-Aid of MEXT, Japan
(\#19540273 and \#21540257).

\end{document}